\documentclass[5p]{elsarticle}

\usepackage{amssymb}
\usepackage{latexsym}
\usepackage{graphics}
\usepackage{graphicx}
\usepackage{amsmath}
\usepackage{booktabs}
\usepackage{color}
\usepackage{epsfig}
\usepackage{lipsum}

\usepackage{lineno,hyperref}
\modulolinenumbers[5]

\journal{Physics Letter B}

\begin{document}
\begin{frontmatter}

\title{Measurement of proton-proton elastic scattering into the Coulomb region  at P$_\text{beam}$ = 2.5, 2.8 and 3.2~GeV/c}

\author[ikp1]{H. Xu\corref{mycorauth}}
\cortext[mycorauth]{Corresponding author}
\ead{h.xu@fz-juelich.de}
\author[ikp1]{Y. Zhou}

\author[ikp1]{U. Bechstedt}
\author[ikp1]{J. B\"oker}
\author[ikp1]{A. Gillitzer}
\author[ikp1]{F. Goldenbaum}
\author[ikp1]{D. Grzonka}
\author[ikp1]{Q. Hu\fnref{present0}}
\fntext[present0]{Present address: Institute of Modern Physics, Chinese Academy of Sciences, Lanzhou, 730000, China}

\author[munster]{A. Khoukaz}
\author[zea]{F. Klehr}
\author[ikp1]{B. Lorentz\fnref{present1}}
\author[ikp1]{D. Prasuhn}
\author[ikp1,bochum]{J. Ritman}
\author[ikp1]{S. Schadmand}
\author[ikp1]{T. Sefzick}
\author[ikp1]{T. Stockmanns}
\author[gwu]{I.I. Strakovsky}

\author[munster]{A. T\"aschner\fnref{present1}}
\fntext[present1]{Present address: GSI Helmholtzzentrum f\"ur Schwerionenforschung GmbH, Darmstadt, 64291, Germany}

\author[ucl]{C. Wilkin}
\author[gwu]{R.L. Workman}

\author[zea]{P. W\"ustner}

\address[ikp1]{Institut f\"ur Kernphysik, Forschungszentrum J\"ulich, 52425 J\"ulich, Germany}
\address[munster]{Institut f\"ur Kernphysik, Universit\"at M\"unster, 48149  M\"unster, Germany}
\address[zea]{Zentralinstitut f\"ur Engineering, Elektronik und Analytik, Forschungszentrum J\"ulich, J\"ulich, 52425, Germany}
\address[bochum]{Ruhr-Universit\"at Bochum, Bochum, 44780, Germany}
\address[gwu]{Data Analysis Center at the Institute for Nuclear Studies, Department of Physics, The George Washington University, Washington, D.C. 20052, USA}
\address[ucl]{Physics and Astronomy Department, UCL, London WC1E 6BT, UK}

\begin{abstract}
The proton--proton elastic differential cross section at very small four momentum transfer squared has been measured at three different incident proton momenta in the range of 2.5 to 3.2~GeV/c by detecting the recoil proton at polar angles close to $90^\circ$. The measurement was performed at COSY with the KOALA detector covering the Coulomb--nuclear interference region. The total cross section $\sigma_\text{tot}$, which has been determined precisely, is consistent with previous measurements. The values of the slope parameter $B$ and the relative real amplitude ratio $\rho$ determined in this experiment alleviate the lack of data in the relevant energy region.  This precise data on $\rho$ might be an important check for a new dispersion analysis. 

\end{abstract}

\end{frontmatter}

For a complete understanding of the proton--proton hadronic interaction it is needed to have experimental data such as the total cross section and the elastic cross section over a wide range of energies and angles. From the optical theorem, the imaginary part of the forward elastic scattering amplitude is related to the total cross section, $\displaystyle Imf_\text{n}(0) = k_\text{cm}\sigma_\text{tot}/4\pi$, where $k_\text{cm}$ is the c.m. momentum of the incident particle. Forward dispersion relations predict the real part of the forward elastic scattering amplitude, $Ref_\text{n}(0)$, as well as $\rho$, which is defined as $\displaystyle \rho\equiv Ref_\text{n}(0)/Imf_\text{n}(0)$. Therefore, the measurement of proton--proton elastic scattering enables a way to determine the related parameters, i.e. $\sigma_\text{tot}$ and $\rho$. This method was employed by many experiments, e.g.~\cite{TOTEM, ATLAS}.  However, it is difficult to extract the value of $\rho$ accurately and the absolute magnitude of the differential cross section can not determine the sign of the real part of the nuclear amplitude. 

Fortunately, the proton--proton elastic scattering is always accompanied by Coulomb scattering. The interference between the Coulomb and the real part of the the nuclear amplitude at small 4--momentum transfer squared, $t$, allows to determine both the sign and the value of $\rho$. The parameter $\rho$ is most sensitive in the Coulomb--Nuclear Interference (CNI) region.  The determination of $\rho$ in the CNI region provides the best chance to gain knowledge about it.  

Traditionally, the proton--proton elastic scattering is described in terms of the Coulomb and nuclear amplitudes, $f_\text{c}$ and $f_\text{n}$. As summarized in~\cite{Block}, at small $t$ one obtains

\begin{equation}
 \frac{\text{d}\sigma}{\text{d}t}=\left|f_{\text{c}}(t)\text{e}^{i\alpha\phi(t)}+f_{\text{n}}(t)\right|^2 = \frac{\text{d}\sigma_{\text{c}}}{\text{d}t}+\frac{\text{d}\sigma_{\text{int}}}{\text{d}t}+\frac{\text{d}\sigma_{\text{n}}}{\text{d}t},
\label{eq:formular}
\end{equation}
where 
\begin{equation}
\frac{\text{d}\sigma_{\text{c}}}{\text{d}t}=\frac{4\pi\alpha^2G^4(t)(\hbar c)^2}{\beta^2t^2},
\label{eq:coulomb}
\end{equation}

\begin{equation}
\frac{\text{d}\sigma_{\text{int}}}{\text{d}t}=-\frac{\alpha\sigma_{\text{tot}}}{\beta\left|t\right|}G^2(t){\text{e}}^{\frac{-B\left|t\right|}{2}}(\rho\cos(\alpha\phi(t))+\sin(\alpha\phi(t))),
\label{eq:inter}
\end{equation}
and
\begin{equation}
\frac{\text{d}\sigma_{\text{n}}}{\text{d}t}=\frac{\sigma_{\text{tot}}^2 (1+\rho^2){\text{e}}^{-B\left|t\right|}}{16\pi(\hbar c)^2}.
\label{eq:hadron}
\end{equation}

Here, $\alpha$ is the fine structure constant. $G(t)$ is the proton dipole form factor conventionally taken as $G(t)=(1+\Delta)^{-2}$, with $\Delta\equiv\left|t\right|/0.71$ (GeV/c)$^2$. $\phi (t)$ is the Coulomb phase, that we take in the form~\cite{West, Cahn},

\begin{equation}
\phi(t)=-ln\left(\frac{B\left|t\right|}{2}\right)-\gamma,
\label{eq:phase}
\end{equation}
where Euler's constant is $\gamma\approx0.577$. There are no free parameters in the Coulomb cross section and so, if this contribution can be determined, it can be used to fix the normalization. The parameters $\sigma_{\text{tot}}$ and $B$ and $\rho$ involved in the parameterization of the nuclear and the interference cross section need to be determined by experiment. 

\begin{figure}[htb]
\centering
\includegraphics[width=1.0\columnwidth]{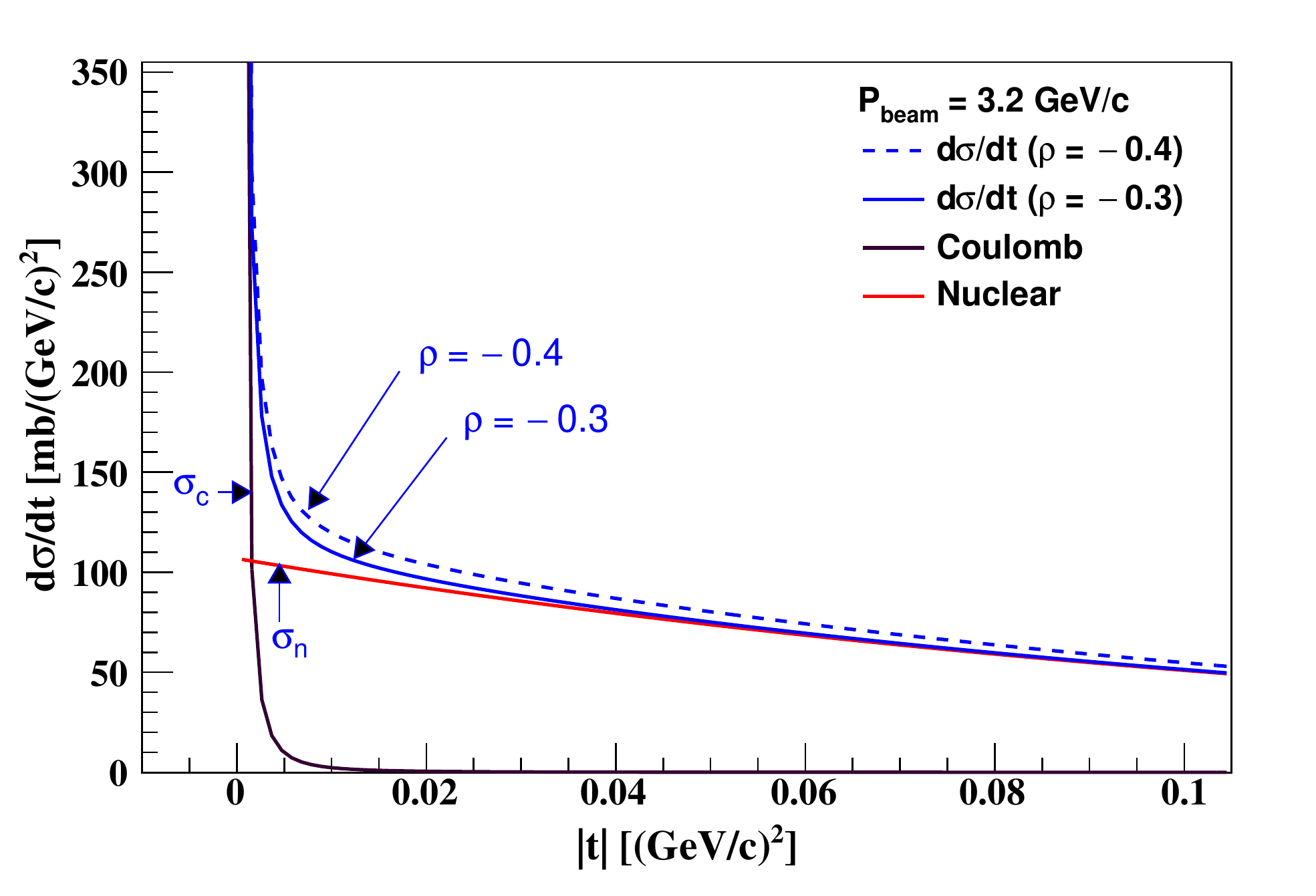}\hspace{5mm}
\caption{Calculated partial and total differential cross section of proton--proton elastic scattering in a wide $t$ range covering the CNI region at 3.2~GeV/c based upon the Eqs.~(\ref{eq:formular})--(\ref{eq:hadron}) with the values of $\sigma_\text{tot}$ and $B$ taken from the world data presented below. The total differential cross section for $\rho = -0.4$ and $\rho = -0.3$ are indicated by the dashed and solid blue lines, respectively.} 
\label{fig:caldsdt}
\end{figure}

It is noted that an early measurement found the contribution of the spin--spin amplitude to the forward elastic scattering~\cite{IKAR1}. Further measurements show that the real parts of the spin--spin amplitude decrease rapidly with increasing beam energy. There is some evidence that the spin-spin amplitude might become negligible already for beam momentum at 1.7~GeV/c~\cite{IKAR} for the nuclear scattering contribution to the cross section. Therefore, the contribution of the spin--spin amplitude to the nuclear differential cross section was neglected in this analysis.

It is necessary to measure the $t$ dependence of the elastic scattering differential cross section over a wide range of $t$ in order to resolve the strong correlation between the fit parameters when they are only measured within a narrow range of $t$~\cite{Koala}. Figure~\ref{fig:caldsdt} indicates the calculated partial and total differential cross section of proton--proton elastic scattering at a beam momentum of 3.2 GeV/c with the parameterization given above. The Coulomb cross section dominates at small $t$. The slope parameter $B$ may be determined from the $t$ dependence far above the CNI region. The parameter $\rho$ can be determined by analysing the $t$--distribution in the CNI region, i.e. the region where $\text{d}\sigma_\text{\text{c}}/\text{d}t\approx\text{d}\sigma_\text{\text{n}}/\text{d}t$. At a beam momentum of 3.2 GeV/c this corresponds to $\left|t\right|\approx0.0015$ (GeV/c)$^2$. A measurement to even lower $t$, where the cross section is dominated by Coulomb scattering, would enable us to determine the luminosity as well as the total cross section by analyzing the characteristic shape of the $t$--distribution over a wide range. One goal of KOALA  is to measure the $|t|$ range of $0.0008-0.1 (\text{GeV/c})^2$, which covers the CNI region. This so called Coulomb normalization method was also pursued by high energy experiments, such as UA4~\cite{UA4} and ATLAS~\cite{ Alfa}.  

 In this report, we present precision measurements of the proton--proton elastic differential cross section at very small $t$ covering the CNI region at three beam momenta of 2.5, 2.8 and 3.2~GeV/c at COSY~\cite{COSY}. The measurements were made by detecting the recoil protons at polar angles close to 90$^\circ$. 

 \begin{figure}
\centering
\includegraphics[width=1.0\columnwidth]{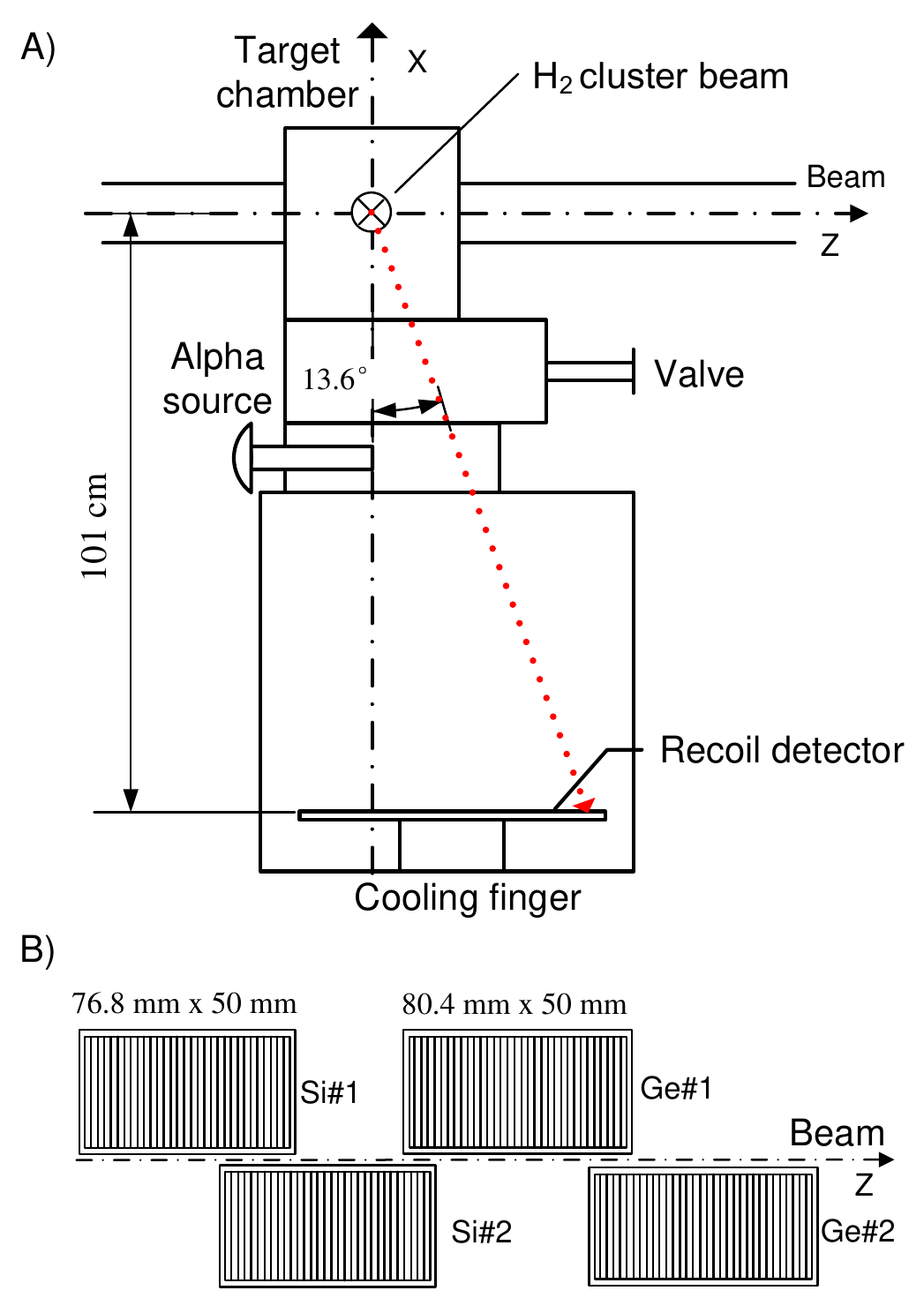}
\caption{A schematic view of the recoil detector system is shown in A). The sensor layout is shown in B), as seen from the interaction region. }
\label{fig:RecoilArm}
\end{figure}

Based on the detection of the recoil proton, the value of $t$ is proportional to the laboratory kinetic energy $T_\text{p}$ of the recoil proton and is related to the recoil angle $\alpha = 90^\circ - \theta$,

\begin{equation}
-t=2m_\text{p}T_\text{p}=\frac{4m^2_\text{p}\sin^2{\alpha}}{(1/\beta^2_\text{cm})-\sin^2{\alpha}},
\label{eq:t}
\end{equation}
where $1/\beta^2_\text{cm}=(E_\text{beam}+m_\text{p})/(E_\text{beam}-m_\text{p}$), and $E_\text{beam}$ and $m_\text{p}$ are the total beam energy and the proton mass, respectively. The recoil measurement technique, which has been used before~\cite{E760,RecoTec}, has several advantages compared to measuring the scattered protons at small forward angles. It is of great importance that the recoil measurement can achieve a larger acceptance than the forward measurement due to the limit of either the beam pipe or the beam emittance. Secondly, it is much easier to distinguish the recoil protons from the non-interacting protons. Thirdly, recoil protons have relatively small energies and their kinetic energy can be precisely detected in solid state detectors. For instance for $|t|=0.1~(\text{GeV/c})^2$, $T_\text{p}=54$~MeV, recoil protons have a range of $<8$ mm in germanium or $<14$ mm in silicon. 
 
The measurements have been made with the recoil setup of the KOALA experiment~\cite{Koala}, which can measure proton--proton elastic scattering in the angular range up to $\alpha=13.6^{\circ}$ for beam momenta between 2.5 to 3.2~GeV/c. The KOALA experiment was proposed to measure antiproton--proton elastic scattering  at HESR~\cite{Xu12}. The KOALA recoil detector system was commissioned at COSY by measuring proton--proton elastic scattering since the recoil particle and the kinematics are identical for both reactions. 

The top part of Figure~\ref{fig:RecoilArm} schematically shows the setup used to measure the recoil protons at COSY. The circulating proton beam in the COSY ring intersects an internal H$_2$ cluster beam target (typical areal density of 10$^{14}$ protons/cm$^2$) to provide a luminosity of about $1\times10^{30}\text{cm}^{-2}\text{s}^{-1}$. The interaction region consists of the intersection of a ~5~mm diameter cylindrical proton beam with an oval shaped hydrogen gas jet with a lateral width of ~10~mm and a thickness of 1--2~mm along the proton beam direction~\cite{Khoukaz}. A sketch of the sensor layout is shown in the bottom of Figure~\ref{fig:RecoilArm}. The detector system has been described elsewhere~\cite{Koala}. As used at COSY, the KOALA recoil detector included two $76.8~\text{mm}\times50~\text{mm}\times1~\text{mm}$ silicon strip sensors that were positioned about 1~m from the target. Each silicon detector has 64 strips with 1.2~mm pitch. In order to measure higher energy protons, two germanium strip detectors with 5 and 11~mm thickness were added. Each has 67 strips and a strip pitch of 1.2~mm.  The four solid state strip detectors were installed on a cold plate viewing the intersection region and covering recoil angles from $\alpha=-1.5^{\circ}$ to $13.6^{\circ}$. A windowless alpha source with three isotopes could be inserted at about 38~cm above the detector surface to allow dedicated energy calibration measurements. 

With the recoil technique the experimental goals consist of precisely measuring the energies of the recoil protons and their relative yield at different recoil angles. In the present experiment the recoil angles typically ranged from $\alpha=0^\circ$ to 13.6$^\circ$ corresponding to $|t| = 0$ to $|t| = 0.1$~(GeV/c)$^2$ at P$_{\text{beam}}=3.2$~GeV/c. The recoil energies are in the range $T_\text{p}$ = 0 to 54~MeV. An alpha source has been used to measure the energy resolution of the silicon and germanium detectors, and was determined to be better than 20 keV and 30 keV (FWHM), respectively. The energy calibration of the detectors was done when no beam was circulating. The detector worked stably during the experiment. The energy calibration measurement for each single strip was reproducible with an uncertainty smaller than 0.05$\%$.   

Prior to the further analysis, an energy clustering algorithm has been implemented in order to reconstruct the proton energy deposited in more than one detector strip. An energy cluster consists of all relevant neighbouring strips, in which the deposited energy is above a strip--dependent threshold. The total energy of each event has been reconstructed based on the energy of the cluster. The overall precision $(\delta E/E)$ of the reconstructed energy is better than 0.3\%. 
\begin{figure}
\centering
\includegraphics[width=1.0\columnwidth]{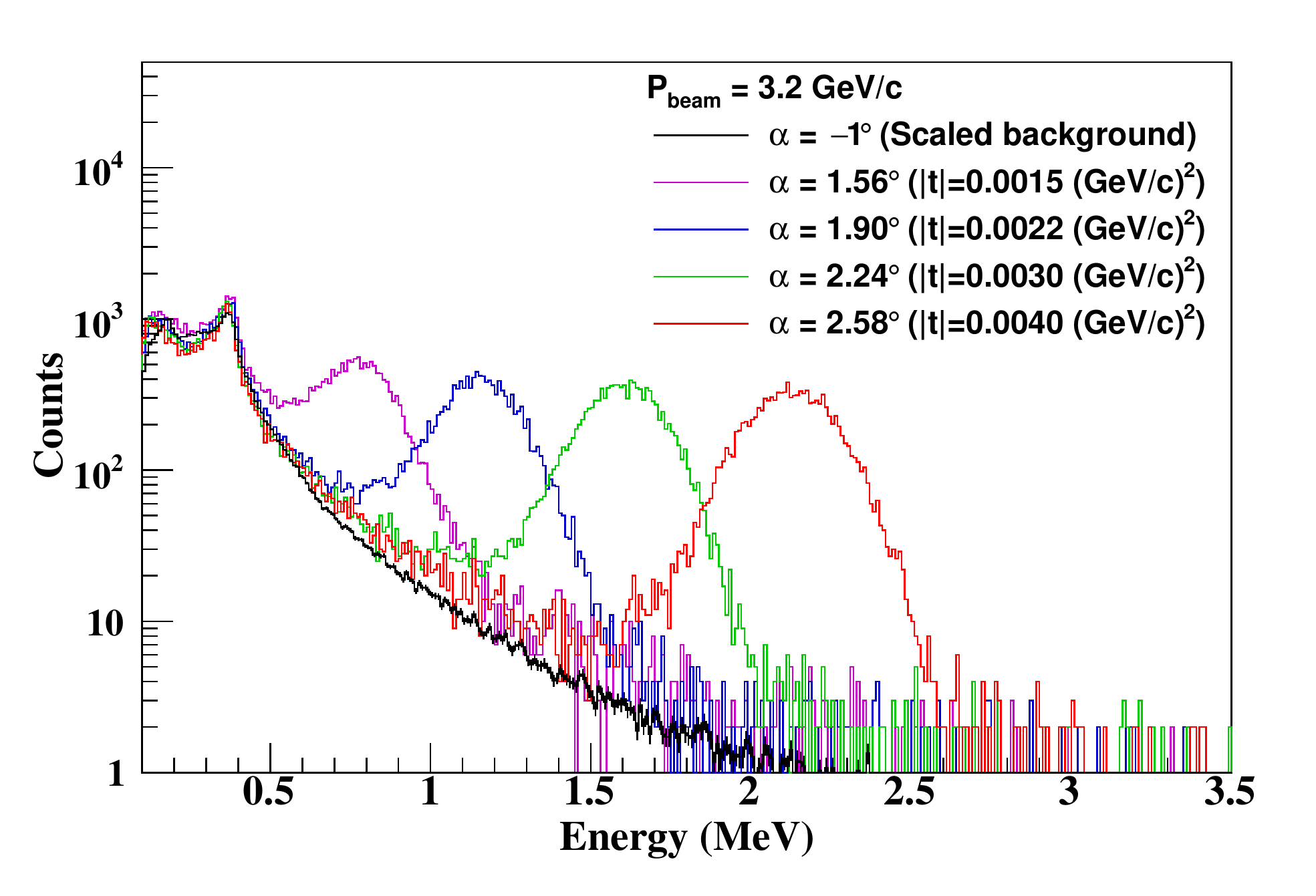}
\caption{The energy distribution of recoil protons at small recoil angles shows the elastic scattering peak above the scaled pure background measured in the unphysical region (black curve).}
\label{fig:BKG}
\end{figure}

Two challenges that have to be faced for gaining high precision of the measurements are the background subtraction in the recoil proton spectra as well as the accurate determination of the mean value of $t$. For $|t|\geqslant$ 0.004~(GeV/c)$^2$, the background influence is very small with a signal to background ratio above 100. For $|t|<$ 0.004~(GeV/c)$^2$ the recoil energy is less than 2 MeV and the recoil peak sits on a background shown by the black distribution in Figure~\ref{fig:BKG}. The background drops quickly and becomes negligible above around 2 MeV. 

By design, there are 20 strips of the first silicon sensor located in the unphysical region with the recoil angle $\alpha=0^\circ$ to $-1.5^\circ$, where there are no elastic events. As depicted in Figure~\ref{fig:BKG}, the two main components of the background consist of the secondary radiation from the surrounding material and minimum ionizing particles (MIP). The MIP background had a higher rate than the elastic events and formed a peak at around 380~keV, consistent with the energy loss of a MIP passing through 1~mm of silicon. It is found that the background distribution on strips covering the unphysical recoil angle were similar to each other and the yield in the MIP peak varied linearly with the solid angle of the strip. This is also true for the strips covering the physical recoil angle. Therefore, the background measured by those 20 strips were summed and subtracted from the single strip yield using a normalization based upon the yield in the MIP peak. The scaling factor for those strips where the MIP peak partially overlapped the elastic scattering was determined by linearly extrapolating the scaling factors along the solid angle, which was observed from the direct normalization of the peaks without overlap. The scaled data background was then subtracted from the energy spectrum for each strip. The remaining peak is taken to be the real recoil protons of elastic scattering and the yield of recoil protons is numerically integrated. The error of the yield was determined based on  the pure statistical error as well as the background error, added in quadrature.  

\begin{figure}
\centering
\includegraphics[width=1.0\columnwidth]{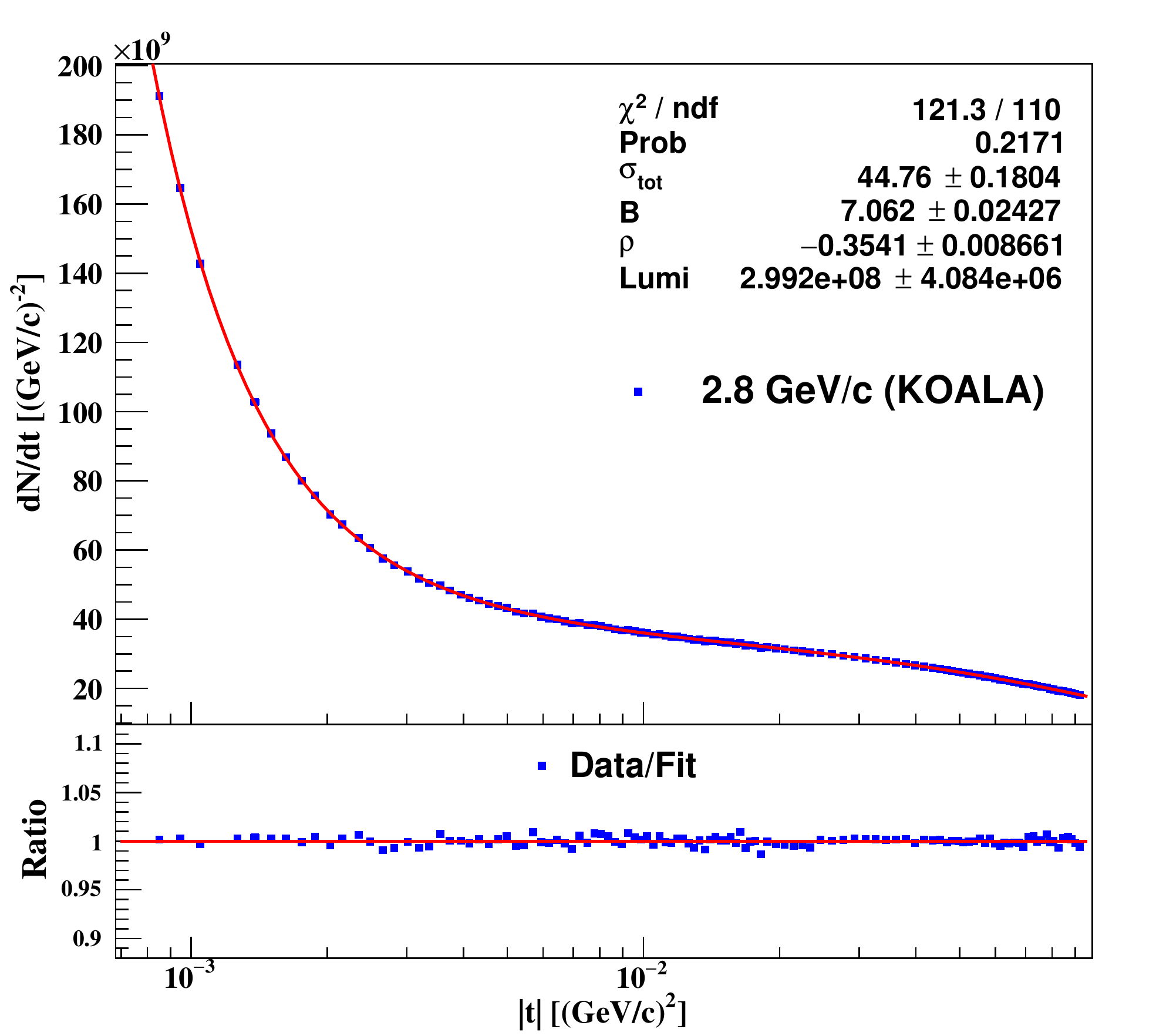}
\caption{The upper part shows the $\text{d}N/\text{d}t$ distribution at 2.8 GeV/c and the corresponding fit parameters. The ratio of the measured data to the fit is shown in the lower part.}
\label{fig:dNdt}
\end{figure}

The determination of the mean value of the momentum transfer, $|t|$, is equivalent to the determination of the mean value of the recoil energy $T_\text{p}$, as indicated in Eq.~(\ref{eq:t}). The $t$--values were calculated based on the mean energy measured by each strip. As a consequence of the high granularity of the detectors, the energy spectrum for a single strip could be well described by a normal distribution. The mean value of the energy was then determined by a Gaussian fit. Due to the background subtraction, the determined $t$ values for $\alpha<1.7^{\circ}$ have larger uncertainties than others. Therefore, $t$ was determined for those strips from the recoil angle based on the geometry of the sensor, i.e. the center of the strip, instead of determining it by using the mean energy. For the other strips, i.e. $\alpha\geqslant1.7^{\circ}$, the $t$ values were accurately determined by the mean energy. The uncertainty of the mean energy determination was added in quadrature with the energy resolution for each strip and taken to be the total error of the determined $t$. 

In parallel to the mean energy determination the detector alignment was implemented. The chamber alignment measurements were done by the COSY crew based upon several permanent benchmark positions in the accelerator tunnel. Except for the finite mechanical installation precision, the hydrogen cluster jet target has also introduced small off--center effects. A software based detector alignment has been performed by comparing the measured proton energy, i.e. the recoil angle, to the expected angle related to the center of each single strip. Based on the alignment, a shift of 1--2~mm relative to the survey has been implemented in order for both determinations to match. This is consistent with the results of the position measurement of the target profile and the detector chamber. With the implementation of the detector alignment, the solid angle for each strip was calculated based on the strip pitch and the distance between the detector surface and the interaction volume. The overall uncertainty of the solid angle determination is smaller than $0.2\%$. 

\begin{figure}
\centering
\includegraphics[width=1.0\columnwidth]{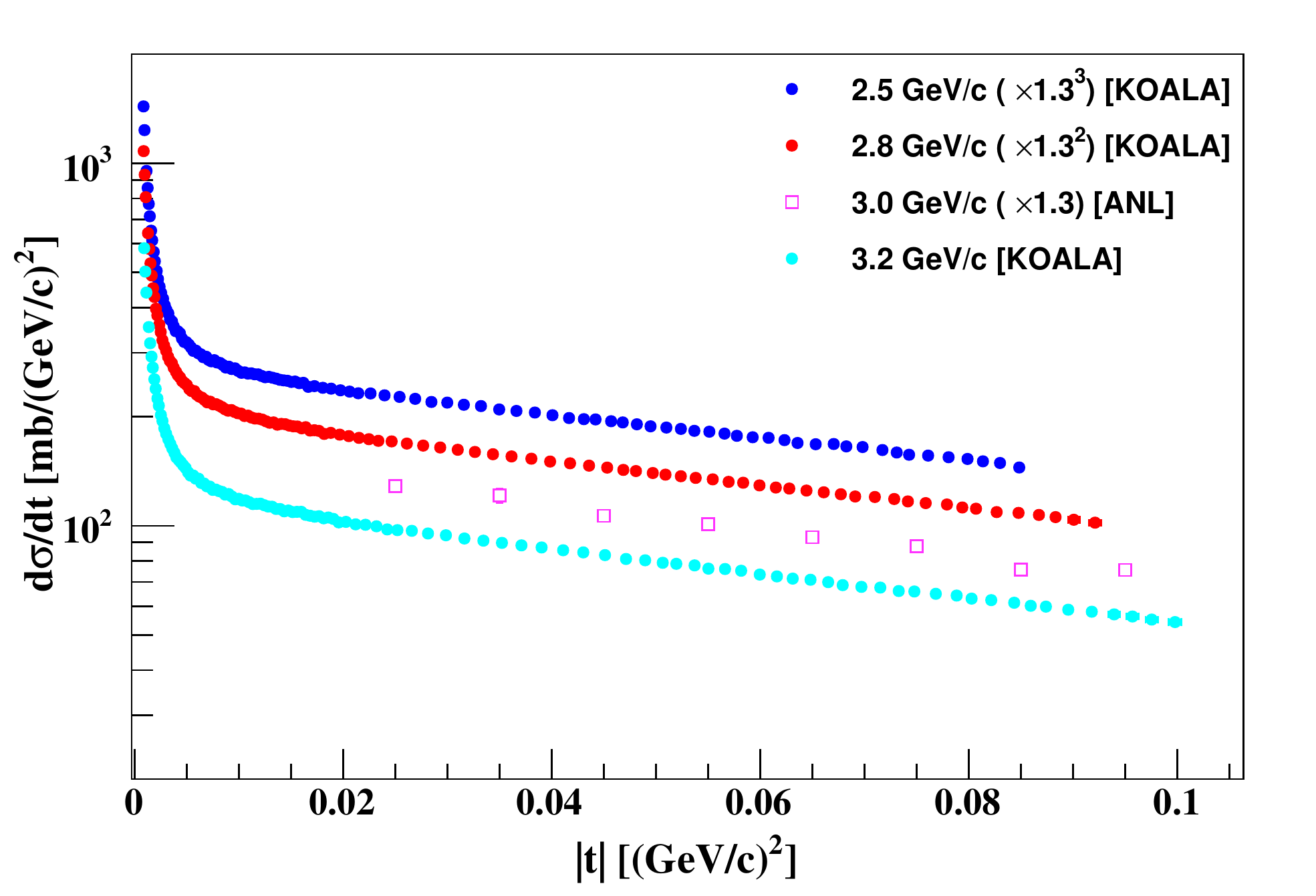}
\caption{The differential cross section distributions measured by KOALA together with a previous measurement of ANL~\cite{ANL}.}
\label{fig:dsigdt}
\end{figure}

The d$N$/\text{d}$t$ distribution was then reconstructed on the basis of the determined $t$, the yield $N$ of elastic events as well as the solid angle d$\Omega$ of each strip, i.e.
\begin{equation}
 \frac{\text{d}N}{\text{d}t}=\frac{\pi}{k_{\text{cm}}^{2}}\frac{\text{d}N}{\text{d}\Omega}.
 \label{eq:dNdO}
 \end{equation}
Since the measurements extended into the Coulomb region, i.e. $|t|<$0.001 (GeV/c)$^2$, all three parameters $\sigma_\text{tot}$, $B$ and $\rho$ as well as the luminosity, $\mathcal{L}$, can be determined from these data. The reconstructed $\text{d}N/\text{d}t$ distribution was fit by the differential cross section formulas in Eqs.~(\ref{eq:formular})--(\ref{eq:hadron}) multiplied by $\mathcal{L}$, i.e. 
\begin{equation}
\frac{\text{d}N}{\text{d}t}=\mathcal{L}\frac{\text{d}\sigma}{\text{d}t}.
\label{eq:dNdt}
\end{equation}

Figure~\ref{fig:dNdt} shows the $\text{d}N/\text{d}t$ distribution at 2.8 GeV/c with a fit marked in red that used Eq.~(\ref{eq:dNdt}). The lower frame of Figure~\ref{fig:dNdt}  shows the ratio of the measured data to the fit result, indicating that the fit very well describes the data. 

\begin{figure}
\centering
\includegraphics[width=1.0\columnwidth]{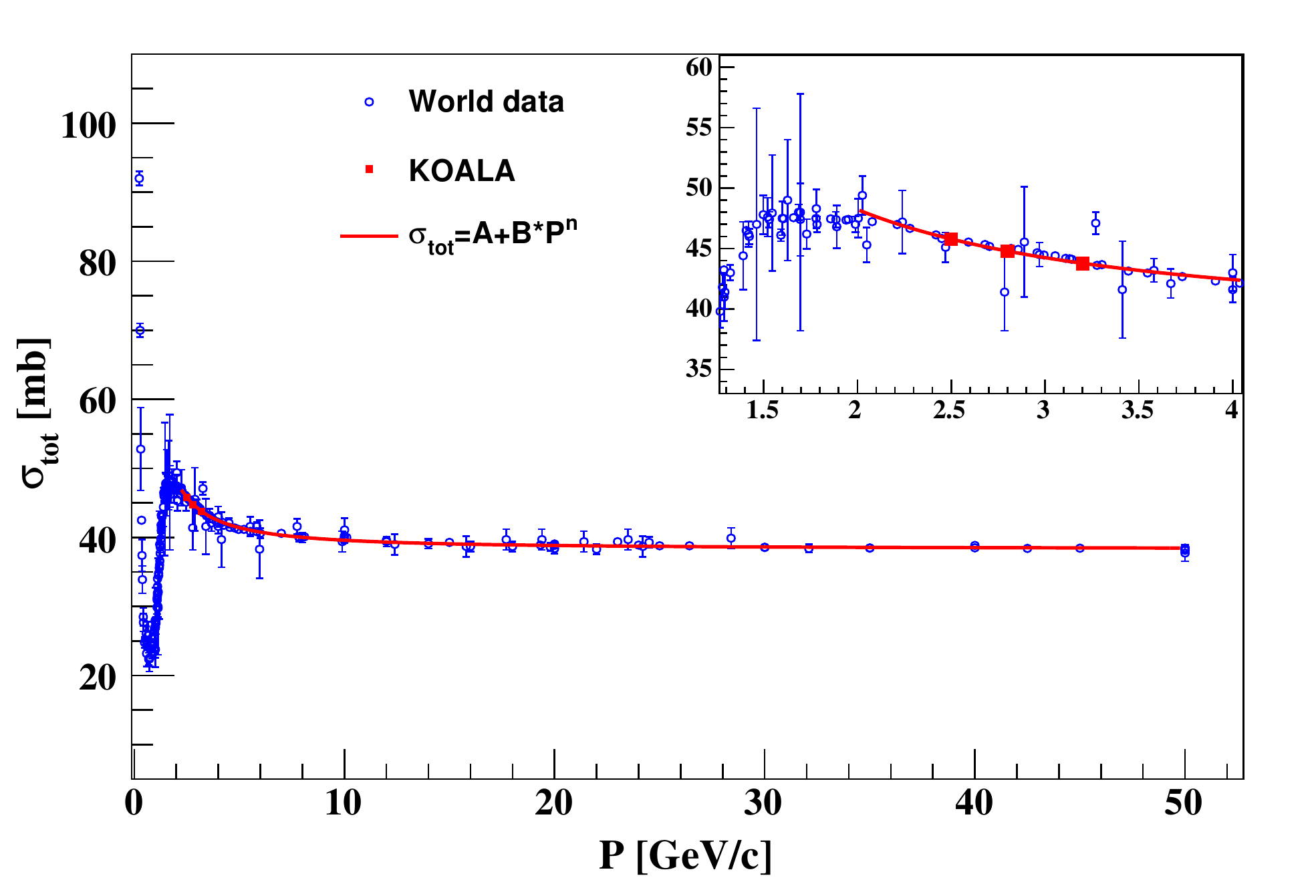}
\caption{The total cross section for this measurement is shown by the red squares in comparison with the world data in blue taken from~\cite{PDG}. The inset displays these results for a narrower range of beam momentum.}
\label{fig:sigtot}
\end{figure}

The differential cross section distributions for all three momenta, as well as a different data sample measured close to our $t$ and P$_{\text{beam}}$ range, have been plotted in Figure~\ref{fig:dsigdt}. The differential cross sections at 2.5, 2.8 and 3.2~GeV/c are presented as solid circles with different colors. Previously measured data at 3~GeV/c~\cite{ANL} in a similar $t$ range are also shown with open squares. For display purposes factors of 1.3, 1.3$^{2}$ and 1.3$^{3}$ have been applied to the data points at 3.0, 2.8 and 2.5~GeV/c, respectively. In contrast to the ANL data the strong rise for $|t|<$0.005 (GeV/c)$^2$ due to the Coulomb interaction is clearly visible in the KOALA data. 

\begin{table}[!ht]
\fontsize{7}{8}\selectfont
\caption{Forward scattering parameters determined for proton--proton elastic scattering. The first error term is the statistical error and the second is the systematic error.}
\label{tab:tab1}\vspace{2mm}\centering
\setlength{\tabcolsep}{0.3\tabcolsep}
 \renewcommand{\arraystretch}{1.5}
\begin{tabular} {*{5}{c}}
	\hline
	P$_\text{lab}$ & $\sigma_\text{tot}$ & $B$ & $\rho$ & $\chi^{2}$/ndf \\ 
	GeV/c & mb & (GeV/c)$^{-2}$ & \\
	\hline
	2.5 & 45.79$\pm$0.27$\pm$0.22 & 6.74$\pm$0.03$\pm$0.02 & --0.314$\pm$0.011$\pm$0.006 & 126.8/109 \\
	2.8 & 44.76$\pm$0.18$\pm$0.33 & 7.06$\pm$0.02$\pm$0.02 & --0.354$\pm$0.010$\pm$0.005 & 121.3/110 \\
	3.2 & 43.76$\pm$0.20$\pm$0.48 & 7.40$\pm$0.02$\pm$0.01 & --0.391$\pm$0.011$\pm$0.005 & 124.0/110 \\
	\hline
\end{tabular}
\end{table}

There are two main sources of systematic error in the determination of the best fit parameters. The first source is related to the background subtraction. As described above, the MIP distribution from the strips with $\alpha<0^{\circ}$ was scaled and then subtracted from the distributions with $\alpha>0^{\circ}$.  The background subtraction introduces an uncertainty of the yield for each individual strip that is smaller than $0.2\%$. Furthermore, the background subtraction could slightly change the shape of the energy peak, which impacts the mean energy, thus $t$, determination. The uncertainty introduced to the $t$ determination caused by the background subtraction is smaller than $0.1\%$. In this analysis, two sets of optimal scaling factors have been implemented for the background subtraction in order to evaluate the systematic error.

The second source of systematic error is the range of the energy spectrum integrated in order to determine the yield of the recoil protons on a given strip. The yield of recoil protons was determined by integrating ranges of width $\pm3\sigma$, $\pm3.5\sigma$ and $\pm4\sigma$ related to the energy peak. The combination of those two sources results in six independent fits for the parameter determination. The RMS of the determined parameters from the six analyses is taken to be the systematic error. These values are presented as the second error term for each parameter in Table~\ref{tab:tab1}.

The elastic scattering parameter with the most existing world data is  the total cross section. These results have been measured by different experiments with various methods, e.g. transmission measurement. Very few results were measured by using the Coulomb normalization method discussed here. Figure~\ref{fig:sigtot} shows the total cross section measured by KOALA together with a compilation of the world data taken from~\cite{PDG}. The KOALA results are in very good agreement with existing data. 

The slope parameter $B$ determined by KOALA is based on Eq.~(\ref{eq:hadron}), which assumes the nuclear differential cross section is described by a pure exponential function. The nuclear elastic differential cross section in the high $t$ region is often described as 

\begin{equation}
\frac{\text{d}\sigma_{\text{n}}}{\text{d}t}=\frac{\text{d}\sigma_{\text{n}}} {\text{d}t}\bigg|_{t=0}\text{exp}(Bt+Ct^{2}).
\label{eq:dSign}
\end{equation}
The value of the nuclear elastic differential cross section extrapolated down to $t=0$, i.e. $\text{d}\sigma_{\text{n}}/\text{d}t|_{t=0}$, is the so called optical point. 

\begin{figure}
\centering
\includegraphics[width=1.0\columnwidth]{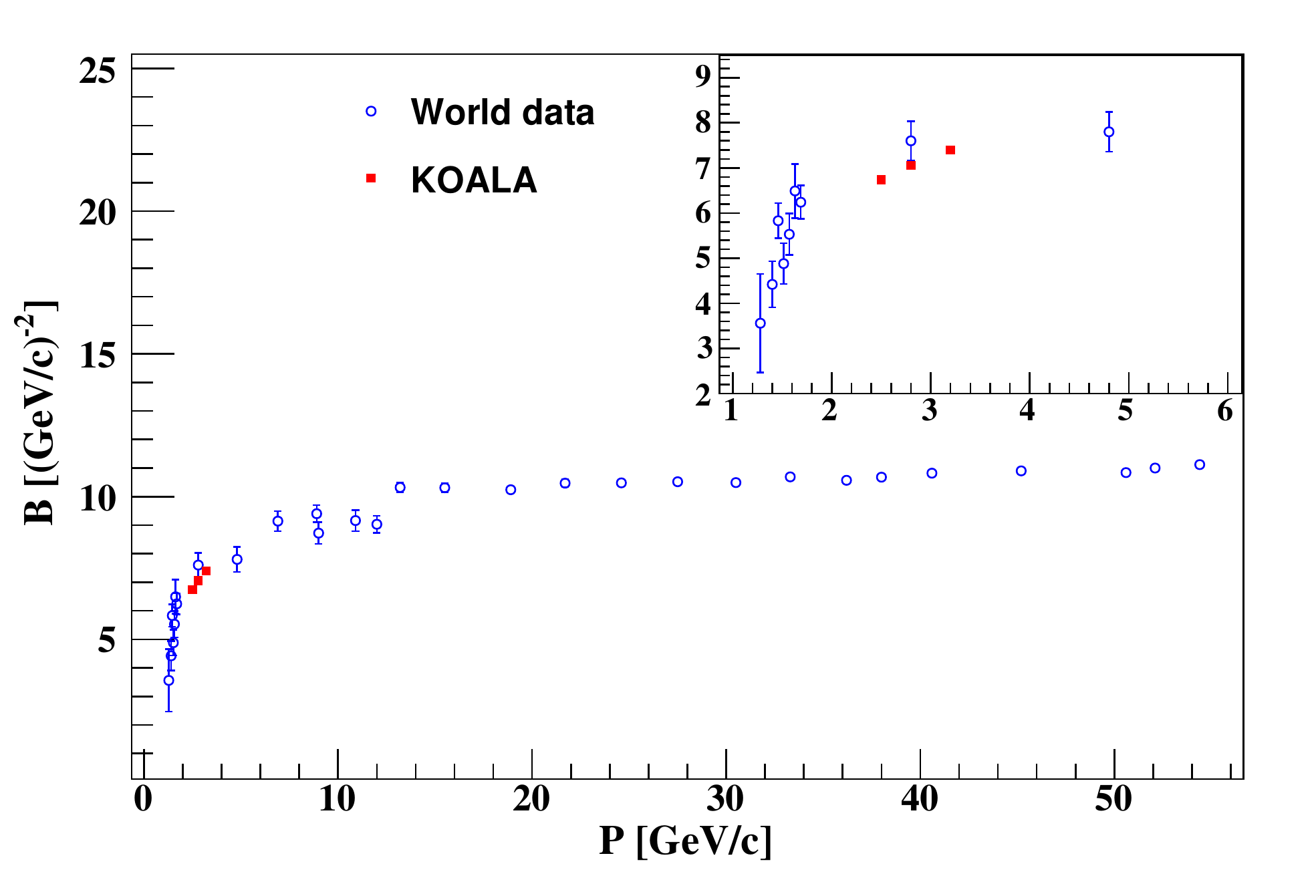}
\caption{The slope $B$ determined by KOALA in red, together with previous measurements in blue~\cite{IKAR, Bez73, Bar73}, for which the same parameterization was applied. The inset displays these results for a narrower range of beam momentum. }
\label{fig:B}
\end{figure}

 In~\cite{Bez73, Bez69} it is shown that for $|t|<$0.12~(GeV/c)$^{2}$ it is possible to take $C$=0. These KOALA measurements have achieved a wide $t$-range, in which the maximum $|t|$ is about 0.1 (GeV/c)$^{2}$, below the cited threshold to change the parameterization. Figure~\ref{fig:B} shows the $B$ values measured by KOALA together with the results from other experiments~\cite{IKAR, Bez73, Bar73}, which used the same parameterization for $B$. The good consistency between the KOALA data and the existing data encourages a more detailed study of how the slope parameter $B$ behaves with the beam energy. 
 
 \begin{figure}
\centering
\includegraphics[width=1.0\columnwidth]{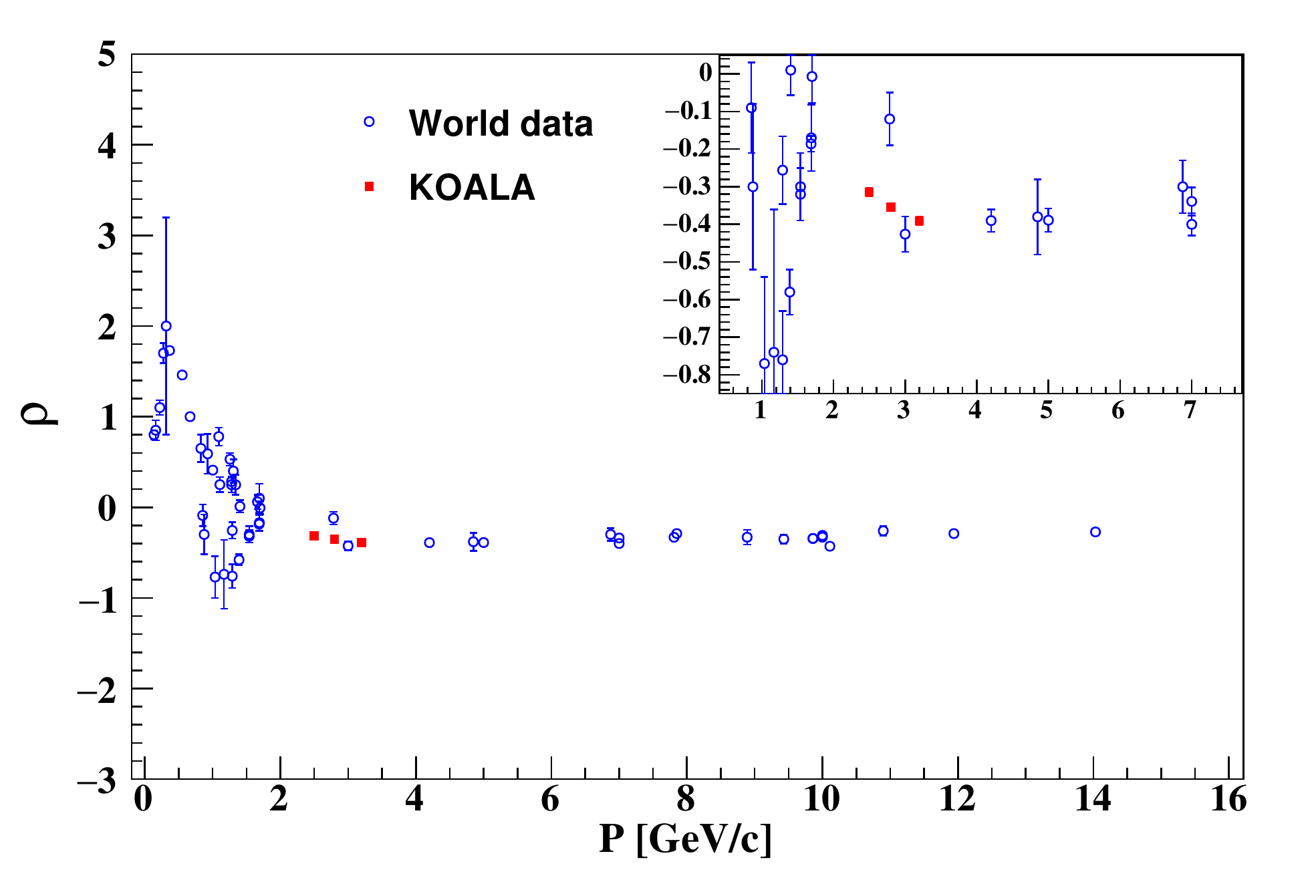}
\caption{The $\rho$ parameter measured by KOALA in red in comparison to previous experiments~\cite{PDG}. The inset displays these results for a narrower range of beam momentum.}
\label{fig:rho}
\end{figure}

\begin{table}[!ht]
\fontsize{9}{9}\selectfont
\caption{Proton--proton elastic scattering parameters when $\sigma_{\text{tot}}$ is held fixed. The statistical error of the parameters is listed.}
\label{tab:tab2}\vspace{2mm}\centering
\setlength{\tabcolsep}{0.6\tabcolsep}
 \renewcommand{\arraystretch}{1.5}
\begin{tabular} {*{5}{c}}
	\hline
	P$_\text{lab}$ & $\sigma_{\text{tot}}$ fixed & $B$ & $\rho$ & $\chi^{2}$/ndf \\ 
	GeV/c & mb & (GeV/c)$^{-2}$ & \\
	\hline
	2.5 & 45.79 & 6.74$\pm$0.03 & --0.314$\pm$0.006 & 126.8/110 \\
	2.8 & 44.79 & 7.06$\pm$0.02 & --0.355$\pm$0.006 & 121.3/111 \\
	3.2 & 43.78 & 7.40$\pm$0.03 & --0.391$\pm$0.008 & 124.0/111 \\
	\hline
\end{tabular}
\end{table}
Benefitting from the excellent detector performance, the lowest values of measured $|t|$ were 0.00083, 0.00085 and 0.0009~(GeV/c)$^{2}$ at 2.5, 2.8 and 3.2~GeV/c, respectively. Consequently both the sign and value of the parameter $\rho$ were able to be determined. The results of $\rho$ measured by KOALA are plotted together with the world data in Figure~\ref{fig:rho}. There are very few data sets of $\rho$~\cite{PDG} to which the KOALA results can be compared in the relevant energy region. 

One challenge while determining the parameter $\rho$ was often the correlation between $\sigma_{\text{tot}}$ and $\rho$ in the narrow $t$ range measured. For a test in this analysis the total cross section was also fixed during the fit in order to check the uncertainty of the determined values of $\rho$.  As shown in Figure~\ref{fig:sigtot}, the world data of the total cross section was fit by the commonly--used parameterization $\sigma_{\text{tot}}=A+B*\text{P}^{n}$~\cite{PDG}, in which $\text{P}$ is the beam momentum, to predict the $\sigma_{\text{tot}}$ at 2.5, 2.8 and 3.2 GeV/c. The corresponding values are given in the second column in Table~\ref{tab:tab2} and they overlap with the KOALA results listed in Table~\ref{tab:tab1}. As a consequence the $\rho$ values determined while $\sigma_{\text{tot}}$ was fixed are quite similar to the case where all parameters are freely varied. In this analysis there is no significant influence from the parameter correlation while determining parameter $\rho$. 

\begin{figure}
\centering
\includegraphics[width=1.0\columnwidth]{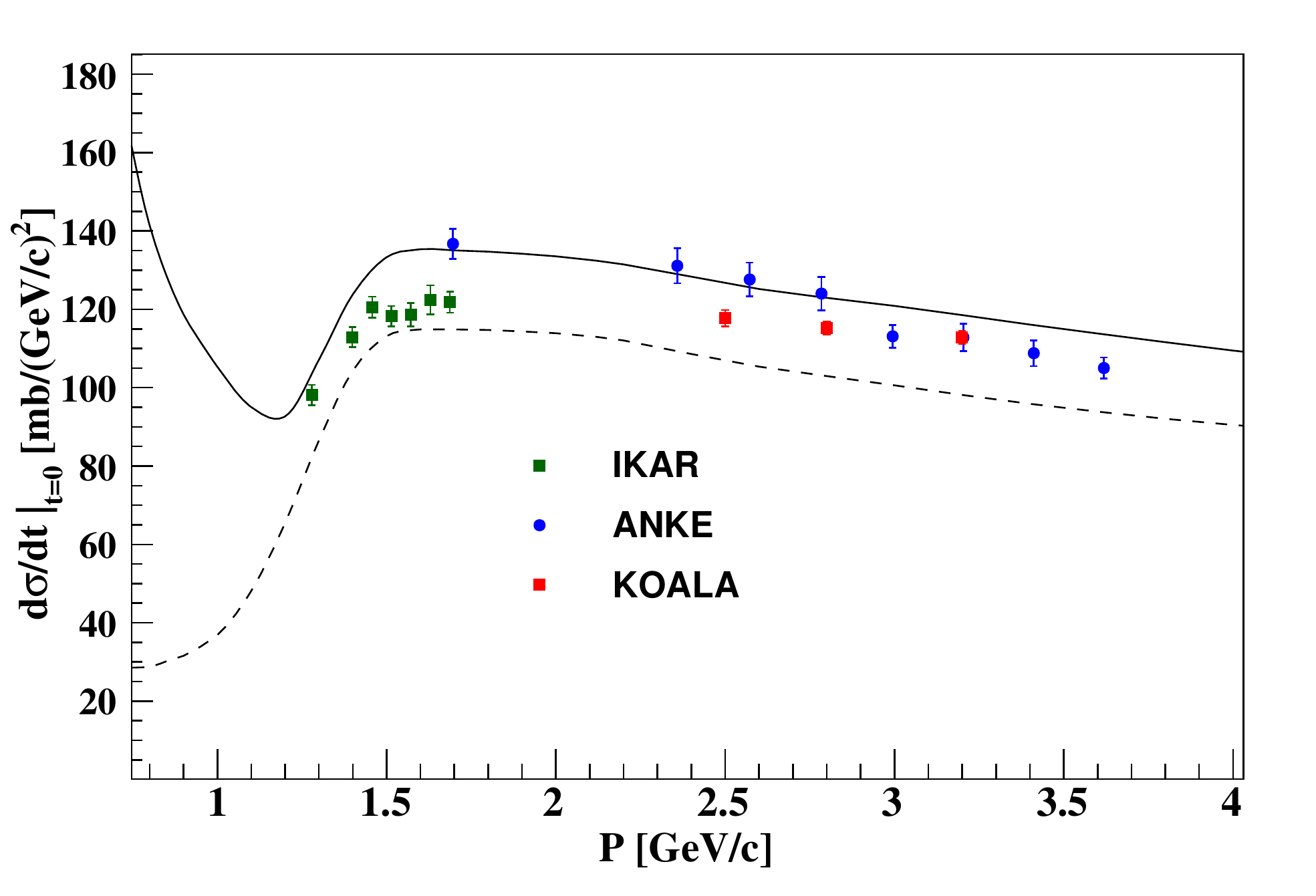}

\caption{The predicted values by Grein and Kroll~\cite{Grein} and the lower limit given by the optical theorem are indicated with solid and dashed line, respectively. The KOALA data are shown with the quoted errors by red squares. The green squares and blue circles are the published IKAR and ANKE values, respectively. The figure is adopted from~\cite{ANKE}.}

\label{fig:OpticalPoint}
\end{figure}

Since the largest $|t|$ reaches $0.1~(\text{GeV/c})^2$, which is far above the CNI region, it enables the extrapolation of the nuclear differential cross section  $\text{d}\sigma_\text{n}/\text{d}t$ down to $t=0$ to determine the optical point with Eq.~(\ref{eq:hadron}) and (\ref{eq:dSign}). Figure~\ref{fig:OpticalPoint} shows the optical point values of the IKAR~\cite{IKAR}, ANKE~\cite{ANKE} as well as KOALA measurements. The solid and dash lines indicate the predicted values~\cite{Grein} and the lower limits~\cite{ANKE} of the optical theorem, respectively. The optical point values for ANKE and KOALA at 3.2~GeV/c nicely match each other. The data at 2.8~GeV/c show a small ($\approx5\%$) discrepancy. However both measurements are compatible when taking the quoted systematic error of 3\% for ANKE and $2.3\%$ for this measurement into account. It is noted that the ANKE points were obtained from SAID fits~\cite{ANKE} instead of a straightforward extrapolation as carried out in this analysis. 
 
In summary, we have measured the proton--proton elastic scattering differential cross section over a wide range of $t$ for three beam momenta using the KOALA recoil detector. The fit parameters $\sigma_{\text{tot}}$, $B$ and  $\rho$ have been determined precisely. The results are consistent with other measurements in the similar energy region. The results contribute to a complete understanding of the hadronic proton--proton interaction. Since there are not many data points on $\rho$ available in this energy range, hopefully the new results gained by KOALA will motivate a new dispersion analysis to predict $\rho$ with even higher precision, which can serve those experiments to measure the total cross section via the optical theorem. 

We are grateful to the COSY crew who installed the device into the COSY ring and provided proton beams. We owe the ANKE collaboration for allowing us to install the recoil detector chamber at their target station. Special thanks for S. Mikirtytchiants, who gave excellent support to find the optimal beam--target overlap. This work was supported in part by the Forschungszentrum J\"ulich COSY-054 FFE under contract number 41808260 and by the U.S. Department of Energy, Office of Science, Office of Nuclear Physics, under awards DE-SC0016582 and DE-SC0016583.

\end{document}